\documentclass[conference]{IEEEtran}
\usepackage{graphicx}
\usepackage{amssymb}
\newcommand{\subst}{\leftarrow}
\newcommand{\img}{\mbox{\rm{i}}}

\newcommand{\bU}{\mbox{\boldmath{$U$}}}

\newcommand{\bm}{\mbox{\boldmath{$m$}}}
\newcommand{\bh}{\mbox{\boldmath{$h$}}}
\newcommand{\bH}{\mbox{\boldmath{$H$}}}

\newcommand{\bD}{\mbox{\boldmath{$D$}}}

\newcommand{\bV}{\mbox{\boldmath{$V$}}}

\newcommand{\bx}{\mbox{\boldmath{$x$}}}

\newcommand{\by}{\mbox{\boldmath{$y$}}}

\newcommand{\bu}{\mbox{\boldmath{$u$}}}

\newcommand{\bzero}{\mbox{\boldmath{$0$}}}

\newcommand{\mR}{\mathbb{R}}
\newcommand{\mN}{\mathbb{N}}

\newcommand{\cu}{\chi_u}
\newcommand{\hcw}{\widehat{\chi}_w}
\newcommand{\hcu}{\widehat{\chi}_u}
\newcommand{\bDelta}{\mbox{\boldmath{$\Delta$}}}
\newcommand{\cA}{{\cal A}}

\ifCLASSINFOpdf
\else
\fi
\begin{document}
\title{An integral formula for large random rectangular matrices 
and its application to analysis of linear vector channels}

\author{\IEEEauthorblockN{Yoshiyuki Kabashima}
\IEEEauthorblockA{Department of Computational Intelligence and Systems Science\\
Tokyo Institute of Technology, Yokohama 226--8502, Japan}
Email: kaba@dis.titech.ac.jp }

%Telephone: (800) 555--1212, Fax: (888) 555--1212}


\maketitle

\begin{abstract}
%\boldmath
A statistical mechanical framework for analyzing 
random linear vector channels is presented in a large system limit. 
The framework is based on the assumptions that the left and right singular 
value bases of the rectangular channel matrix $\bH$ 
are generated independently 
from uniform distributions over Haar measures
and the eigenvalues of $\bH^{\rm T}\bH$ asymptotically 
follow a certain specific distribution. 
These assumptions make it possible to characterize 
the communication performance of the channel utilizing an 
integral formula with respect to $\bH$, which is analogous to 
the one introduced by Marinari {\em et. al. }
in {\em J. Phys. A} {\bf 27}, 7647 (1994) for large random 
square (symmetric) matrices. 
A computationally feasible algorithm for approximately decoding
received signals based on the integral formula is also provided. 
\end{abstract}


% \fi
\IEEEpeerreviewmaketitle

\section{Introduction}
In a general scenario for linear vector channels, 
multiple messages are transmitted to the receiver, 
being linearly transformed to multiple output signals by a 
random matrix and degraded by channel noises. 
This yields a complicated dependence 
 on message variables, which ensures that the problem of inferring   
the transmitted messages from the received output signals is non-trivial. 
In general, inference problems of this kind can be mapped to virtual 
magnetic systems governed by random interactions \cite{Nishimori2001}. 
This similarity has promoted a sequence of statistical mechanical analyses of  
linear vector channels in a large system limit 
from the beginning of this century \cite{Tanaka2001,Kabashima2003,
Muller2003,Moustakas2003,GuoVerdu2005,NeirottiSaad2005,MontanariTse2006}. 

In the simplest analysis, each entry of the channel matrix is regarded as
an independent and identically distributed (IID) random variable. 
However, such a treatment is not necessarily adequate for describing 
realistic systems, in which non-negligible statistical correlations 
across the matrix entries are created by spatial/time proximity 
of messages/antennas or matrix design
for enhancement of communication performance. 
Therefore, the development of methodologies that can deal with 
correlations in the channel matrix is of great importance to research 
in the area of linear vector channels.

It is intended that the present article should 
contribute such a methodology for application 
to these communication channels. 
More precisely, we develop a statistical mechanical framework for analyzing 
linear vector channels so that the influence of the correlations 
across the matrix entries can be taken into account. 
The developed framework is applicable not only to
Gaussian channels of Gaussian inputs \cite{TulinoVerdu2004},
but also general memory-less channels of continuous/discrete inputs,
which are characterized by a factorizable prior distribution.

This article is organized as follows: In the next section, 
the model of linear vector channels that we focus on herein 
is defined. In section III, which is the main part of the current 
article, an integral formula with respect to large random rectangular 
matrices is introduced. A scheme to assess the performance of the 
linear vector channel and a computationally feasible approximate 
decoding algorithm are developed on the basis of this formula. 
The utility of the developed schemes is examined in section IV by 
application to an example system. The final section summarizes the present study's findings.

\section{Model definition}
For simplicity, we here assume that all the variables
relevant to the communication are real; but extending the 
following framework to complex variables is straightforward 
\cite{TakedaHatabuKabashima2007}. 
Let us suppose a linear vector channel
in which an input message vector of $K$ components, 
$\bx=(x_k)$, is linearly transformed to an $M$ dimensional  
sequence, $\bDelta=(\Delta_\mu)$, by a $K \times N$ channel matrix, 
$\bH=(H_{\mu k})$, as $\bDelta=\bH \bx$. 
For generality and simplicity, 
we assume a general memory-less channel, which implies 
that an $N$ dimensional output signal vector, $\by=(y_\mu)$, follows
a certain factorizable conditional distribution as
\begin{eqnarray}
P(\by|\bx;\bH)=P(\by|\bH\bx)=\prod_{\mu=1}^N P(y_\mu|\Delta_\mu). 
\label{channel}
\end{eqnarray}
In addition, we assume a factorizable prior distribution 
\begin{eqnarray}
P(\bx)=\prod_{k=1}^K P(x_k), 
\label{prior}
\end{eqnarray}
for $\bx$, which may be continuous or discrete.

An expression of the singular value decomposition of $\bH$
\begin{eqnarray}
\bH=\bU \bD \bV^{\rm T}, 
\label{singular}
\end{eqnarray}
is the basis of our framework. Here, the superscript ${\rm T}$ denotes the transpose of  
the matrix to which it is attached, 
$\bD={\rm diag}(d_k)$ is an $N \times K$ diagonal 
matrix composed of the singular values $d_k$ $(k=1,2,\ldots, 
{\rm min}(N,K)$), where
${\rm min}(N,K)$ denotes the lesser value of $N$ and $K$. 
The values $d_k$ are linked to the eigenvalues 
of $\bH^{\rm T}\bH$, $\lambda_k$, as
$\lambda_k=d_k^2$ for $k=1,2,\ldots,{\rm min}(N,K)$. 
$\bU$ and $\bV$ are orthogonal matrices 
of order $N \times N$ and $K \times K$, respectively. 
In order to handle correlations in $\bH$ analytically, we assume that 
$\bU$ and $\bV$ are independently generated from 
uniform distributions of the Haar measures of $N \times N$ and 
$K \times K$ orthogonal matrices, respectively, and that 
the empirical eigenvalue distribution of $\bH^{\rm T}\bH$, 
$K^{-1} \sum_{k=1}^K \delta(\lambda-\lambda_k)
=(1-{\rm min}(N,K)/K)\delta(\lambda)+K^{-1}
\sum_{k=1}^K \delta(\lambda-d_k^2)$ converges to 
a certain specific distribution $\rho(\lambda)$
in the limit as $N$ and $K$ tend to infinity while keeping the 
load $\beta=K/N \sim O(1)$. Controlling $\rho(\lambda)$ allows us to 
express various second-order correlations in $\bH$.

\section{Analysis}
\subsection{An integral formula for large random rectangular matrices}
With knowledge of $\bH$, the receiver
decodes $\by$ in order to infer $\bx$, which is performed 
on the basis of the Bayes formula
\begin{eqnarray}
P(\bx|\by;\bH)=
\frac{
\prod_{\mu=1}^N P(y_\mu|\Delta_\mu)\prod_{k=1}^K P(x_k)}
{P(\by;\bH)}. 
\label{posterior}
\end{eqnarray}
Here, the probability
\begin{eqnarray}
P(\by;\bH)=\mathop{\rm Tr}_{\bx}\prod_{\mu=1}^N P(y_\mu|\Delta_\mu)
\prod_{k=1}^K P(x_k), 
\label{partition_function}
\end{eqnarray}
expresses the marginal probability with respect to $\by$,
where $\mathop{\rm Tr}_{\bx}$ denotes summation or integration over the 
all possible states of $\bx$.
%depending on whether $\bx$ is discrete or continuous respectively. 
Eq. (\ref{partition_function}) also serves as the partition function 
concerning the message vector $\bx$ in statistical mechanics. 

Let us examine statistical properties of 
Eq. (\ref{partition_function}) 
prior to  analyzing Eq. (\ref{posterior}). 
The expression 
\begin{eqnarray}
P(\by;\bH)=\mathop{\rm Tr}_{\bu,\bx}
\exp \left (\img \bu^{\rm T}\bH \bx \right )
\prod_{\mu=1}^N \widehat{P}(y_\mu|u_\mu)
\prod_{k=1}^K P(x_k), 
\label{conversion}
\end{eqnarray}
is useful for this purpose, where $\img=\sqrt{-1}$, 
$\bu=(u_\mu)$ and $\widehat{P}(y_\mu|u_\mu)=(2 \pi)^{-1} 
\int d\Delta_\mu \exp \left (-\img u_\mu \Delta_\mu \right )
P(y_\mu|\Delta_\mu)$ denotes the Fourier transformation 
of likelihood $P(y_\mu|\Delta_\mu)$. 
We substitute $\bH$ in Eq. (\ref{conversion}) 
by Eq. (\ref{singular})  and 
take an average with respect to $\bU$ and $\bV$. 
For this assessment, it is noteworthy that for any fixed
set of $\bu$ and $\bx$, $\widetilde{\bu}=\bU^{\rm T} \bu$ and 
$\widetilde{\bx}=\bV^{\rm T} \bx$ behave as continuous 
random variables that satisfy strict constraints
$N^{-1} |\widetilde{\bu}|^2 
=N^{-1} |\bu|^2=T_u$ and
$K^{-1} |\widetilde{\bx}|^2 
=K^{-1} |\bx|^2=T_x$. 
In the %large system 
limit $N,K \to \infty$ keeping $\beta=K/N \sim O(1)$,
which we will hereafter assume if necessary, this yields an expression
\begin{eqnarray}
\frac{1}{N}\ln \left (
\overline{\exp \left (\img \bu^{\rm T} \bH \bx \right )}
\right )=F(T_x,T_u), 
\label{Fxy}
\end{eqnarray}
where $\overline{\cdots}$ denotes the average with respect to 
$\bU$ and $\bV$, and 
\begin{eqnarray}
&&F(\xi,\eta)=\mathop{\rm Extr}_{\Lambda_\xi,\Lambda_\eta}
\left \{ 
-\frac{\beta }{2 }
\left \langle \ln (\Lambda_\xi \Lambda_\eta + \lambda)
\right \rangle_{\rho}
-\frac{1-\beta}{2} \ln \Lambda_\xi \right . \cr
&& \phantom{F(\xi,\eta)}
\left . +\frac{\beta \Lambda_\xi \xi}{2}+\frac{\Lambda_\eta \eta}{2} 
\right \} \! - \! \frac{\beta}{2}\ln \xi 
\! - \! \frac{1}{2}\ln \eta \! - \!
\frac{1+\beta}{2},
\label{F_func}
\end{eqnarray}
where $\left \langle \cdots \right \rangle_\rho$ denotes the average with respect to $\rho(\lambda)$, while 
${\rm Extr}_\theta\left \{ \cdots \right \}$ represents 
extremization with respect to $\theta$ \cite{Kabashima2007}.
This formula is analogous to the one known for ensembles of 
random square (symmetric) matrices 
\cite{MarinariParisiRitort1994,ParisiPotters1995,ItzyksonZuber1980},
which is closely related to the $R$-transformation 
developed in free probability theory
\cite{VoiculescuDykemaNica1992,TulinoVerdu2004,Tanaka2007}. 
Several integral formulae for large random matrices 
related to Eq. (\ref{Fxy}), but for different 
large system 
limits, are presented in \cite{ZinjustinZuber2003}. 

Eq. (\ref{Fxy}) implies
\begin{eqnarray}
&& \frac{1}{N}\ln \left (
\mathop{\rm Tr}_{\by} 
\overline{P(\by;\bH)} \right ) \cr
&&=\mathop{\rm Extr}_{T_x,T_u} \left \{
F(T_x,T_u) +\beta A_x(T_x) +A_u(T_u)\right \}, 
\label{annealed}
\end{eqnarray}
where
$
A_x(T_x)\!=\! \mathop{\rm Extr}_{\widehat{T}_x}\left \{
\widehat{T}_x T_x/2 \! +\! \ln 
\left ( \! \mathop{\rm Tr}_x P(x) e^{-\widehat{T}_x x^2/2}
\! \right )
\! \right \}
$
and
$
A_u(T_u)\!=\!\mathop{\rm Extr}_{\widehat{T}_u}
\!\! \left \{ \! 
\widehat{T}_u T_u/2 \!+\! \ln \!
\left ( \! \mathop{\rm Tr}_{y,u}
\! \widehat{P}(y|u) e^{-\widehat{T}_u u^2/2}
\! \right )\! 
\right \}. 
$
The normalization constraint $\mathop{\rm Tr}_{\by} \overline{P(\by;\bH)}=1$, 
in conjunction with the extremization in Eq. (\ref{annealed}), yields
$T_x=\mathop{\rm Tr}_{x}x^2 P(x)$, $\widehat{T}_x=0$, $T_u=0$ and 
$\widehat{T}_u=\beta \left \langle \lambda \right \rangle_\rho T_x $. 
The physical implication of these results is that components of $\bDelta=\bH \bx$ 
behave as IID Gaussian variables of zero mean and variance 
$\widehat{T}_u$ in the large system limit when $\bx$ is drawn 
from Eq. (\ref{prior}), while $\bU$ and $\bV$ are independently 
generated from the Haar measures. 

\subsection{Performance assessment}
Now, we are ready to analyze the typical communication performance of the 
current channel model. 
This is performed by assessing the 
typical mutual information (per output signal) 
between $\bx$ and $\by$, ${I}(X,Y)$, based on Eqs. (\ref{posterior}) 
and (\ref{partition_function}) as
\begin{eqnarray}
&&{I}(X,Y)=
\frac{1}{N}
\mathop{\rm Tr}_{\by,\bx}\overline{P(\by|\bx;\bH)P(\bx)
\ln \left (\frac{P(\by|\bx;\bH)}{P(\by;\bH)} \right )
} \cr
&&={\cal F}+\mathop{\rm Tr}_{y} 
\int Dz P\left (y|\sqrt{\widehat{T}_u}z \right )
\ln P\left (y|\sqrt{\widehat{T}_u}z \right ), 
\label{mutual_infor}
\end{eqnarray}
where
\begin{eqnarray}
{\cal F}=-\frac{1}{N} \mathop{\rm Tr}_{\by}
\overline{P(\by;\bH)\ln P(\by;\bH)}, 
\label{free_energy}
\end{eqnarray}
represents the conditional entropy of $\by$, and 
serves as the average free energy with respect to $\bx$. 
%The operator 
$Dz=(2 \pi)^{-1/2} dz e^{-z^2/2}$ denotes the Gaussian measure. 
The statistical properties of $\bDelta$ evaluated in the last paragraph
are employed to assess the second term 
on the right-hand side of the last line of Eq. (\ref{mutual_infor}). 

${\cal F}$ can be evaluated by means of the replica method. 
Namely, we evaluate the $n$-th moment of the partition 
function $P(\by;\bH)$ for $n \in \mN$ as
\begin{eqnarray}
&&\mathop{\rm Tr}_{\by} \overline{P^{n+1}(\by;\bH) } 
=\mathop{\rm Tr}_{\by,\{\bx^a\},\{\bu^a\}}
\overline{\exp \left (
\img \sum_{a=1}^n (\bu^a)^{\rm T}\bH \bx^a \right )} \cr
&& 
\times \prod_{a=1}^{n+1} \prod_{\mu=1}^N \widehat{P}(y_\mu|u_\mu^a) 
\times \prod_{a=1}^{n+1} \prod_{k=1}^K P(x_k^a), 
\label{moments}
\end{eqnarray}
and assess ${\cal F}$ as
\begin{eqnarray}
{\cal F}=-\lim_{n \to 0}\frac{\partial}{\partial n}
\frac{1}{N} 
\ln \left (\mathop{\rm Tr}_{\by} \overline{P^{n+1}(\by;\bH) } \right ), 
\label{replica_method}
\end{eqnarray}
analytically continuing expressions obtained for Eq. (\ref{moments})
from $n \in \mN$ to $n \in \mR$. 
Here, $\{\bx^a\}$ denotes a set of $n+1$ replicated vectors 
$\bx^0, \bx^1,\cdots, \bx^n$,  with $\{\bu^a\}$ defined similarly. 

Eq. (\ref{replica_method}) is generally expressed 
using $F(\xi,\eta)$, and the derivation of the expression can be found 
in \cite{Kabashima2007}. 
In particular, the expression obtained under the replica 
symmetric ansatz, which is believed to be correct 
for the current case since the inference is performed
on the basis of the correct posterior (\ref{posterior}) 
\cite{Nishimori2002}, 
is given in a compact form as 
\begin{eqnarray}
{\cal F}=-\mathop{\rm Extr}_{q_x,q_u}
\left \{\cA_{xu}(q_x,q_u)+\beta \cA_x (q_x)+\cA_u(q_u) \right \}, 
\label{replica_free}
\end{eqnarray}
where 
\begin{eqnarray}
\cA_{xu}(q_x,q_u)=F(T_x-q_x,q_u)+\frac{\widehat{T}_u q_u}{2}, 
\label{A0}
\end{eqnarray}
\begin{eqnarray}
\cA_x(q_x)=\mathop{\rm Extr}_{\widehat{q}_x}
\left \{-\frac{\widehat{q}_x q_x}{2} 
+\int Dz {\cal P}(z;\widehat{q}_x) \ln {\cal P}(z;\widehat{q}_x)
\right \}
\label{Ax}, 
\end{eqnarray}
and
\begin{eqnarray}
&&\cA_u(q_u)=\mathop{\rm Extr}_{\widehat{q}_u}
\left \{-\frac{\widehat{q}_u q_u}{2} \right . \cr
&& \left . 
\phantom{\cA_x=}
+\mathop{\rm Tr}_{y}
\int Dz {\cal P}(y|z;\widehat{q}_u ) \ln {\cal P}(y|z;\widehat{q}_u)
\right \}, 
\label{Au}
\end{eqnarray}
in which ${\cal P}(z;\widehat{q}_x)=
\mathop{\rm Tr}_x P(x) 
e^{-\widehat{q}_xx^2/2+\sqrt{\widehat{q}_x}z x}$
and ${\cal P}(y|z;\widehat{q}_u)=
\int Ds P\left (y|\sqrt{\widehat{T}_u -\widehat{q}_u }s
+\sqrt{\widehat{q}_u} z \right )$. 
The $q_x$ and $q_u$ determined by Eq. (\ref{replica_free})
represent $K^{-1} \left [ |\left \langle \bx \right \rangle |^2 \right ]$
and $-N^{-1} \left [ |\left \langle \bu \right \rangle |^2 \right ]$, 
respectively, where $\left \langle \cdots \right \rangle $
denotes averaging over the posterior distribution (\ref{posterior}) 
while $\left [ \cdots  \right ] $ indicates the 
average with respect to $\by$, $\bU$ and $\bV$. 
These averages, $\left \langle \cdots \right \rangle $ and 
$\left [ \cdots \right ] $,
correspond to the thermal and quenched averages in 
statistical mechanics, respectively. 
The quantities $\widehat{q}_x$ and $\widehat{q}_u$  appearing in Eqs. 
(\ref{Ax}) and (\ref{Au}) can be used for assessing 
performance measures other than Eq. (\ref{mutual_infor}), 
such as the mean square error (MSE) and the bit error rate (BER). 

\subsection{Computationally feasible approximate decoding}
Let us suppose a situation which requires evaluation of the 
posterior average
\begin{eqnarray}
\bm_x=\mathop{\rm Tr}_{\bx} \bx P(\bx|\by;\bH), 
\label{postav}
\end{eqnarray}
 where $\bm_x=(m_{xk})$, with similar
notation used for other vectors below. 
Eq. (\ref{postav}) serves as the estimator that 
minimizes the MSE in general, and can be used to minimize 
the BER for binary messages. Exact assessment 
of such averages is, however, computationally difficult 
for large systems, which motivates us to develop computationally 
feasible approximation algorithms 
\cite{VaranasiAazhang1990,Kabashima2003,TanakaOkada2005}. 
A generalized Gibbs free energy 
\begin{eqnarray}
&&
\widetilde{\Phi}(\bm_x,\bm_u;l)=\mathop{\rm Extr}_{\bh_x,\bh_u}
\left \{ \bh_x \cdot \bm_x +\bh_u \cdot \bm_u \right . \cr
&& \left .
\phantom{\widetilde{\Phi}(\bm_x,\bm_u;l)=}
-\ln \left (Z(\bh_x,\bh_u;l)\right)
\right \}, 
\label{gen_gibbs_FE}
\end{eqnarray}
where $Z(\bh_x,\bh_u;l) = \mathop{\rm Tr}_{\bx,\bu} 
\prod_{\mu=1}^N \widehat{P}(y_\mu|u_\mu)
\times 
\prod_{k=1}^K P(x_k)$ 
$\times$ 
$\exp \left (
\bh_x \cdot \bx+\bh_u \cdot ({\rm i}\bu)+
({\rm i} \bu)^{\rm T}(l\bH) \bx \right )$, 
offers a useful basis for this purpose 
since Eq. (\ref{postav}) is characterized as 
the unique saddle point of Eq. (\ref{gen_gibbs_FE})
for $l=1$ \cite{Plefka1982,OpperWinther2005}. 
\begin{figure}[t]
\small
{\bf {MPforPerceptron}}$\{$
\begin{eqnarray}
&&\mbox{Perform {\bf {Initialization}}}; \cr
&& \mbox{Iterate {\bf {H-Step}} and {\bf {V-Step}
} alternately sufficient times;}
\nonumber
\end{eqnarray}
$\}$\\
%\end{quote}
%\begin{quote}
{\bf {Initialization}}$\{$
\begin{eqnarray}
&&\chi_x \subst \frac{1}{K} \sum_{k=1}^K x_k^2 P(x_k); 
\quad \widehat{\chi}_x \subst 0; 
\quad \Lambda_x \subst \frac{1}{\chi_x}-\widehat{\chi}_x;\cr
&&m_{xk} \subst \mathop{\rm Tr}_{x_k}x_k P(x_k) \qquad (k=1,2,\ldots,K); \cr
&&\bh_u \subst \bH \bm_x;\quad \bm_u \subst \bzero;
\nonumber
\end{eqnarray}
$\}$\\
{\bf {H-Step}}$\{$
\begin{eqnarray}
&&\mbox{Search $(\cu,\Lambda_u)$ for given $(\chi_x,\Lambda_x)$ 
to satisfy conditions} \cr
&& \quad\chi_x\!=\!\left \langle \frac{\Lambda_u}{\Lambda_x \Lambda_u+\lambda} 
\right \rangle_\rho \mbox{and } 
\cu\!=\!\frac{1-\beta}{\Lambda_u}
\!+\!\left \langle \frac{\beta \Lambda_x}{\Lambda_x \Lambda_u+\lambda} 
\right \rangle_\rho; \cr
&&\hcu\subst \frac{1}{\cu}-\Lambda_u; \cr
&& \bh_u \subst \bh_u-\hcu \bm_u; \cr
&& m_{u\mu} \subst \frac{\partial}{\partial h_{u\mu}}
\ln \left (\int Dx
P(y_\mu|\sqrt{\hcu}x+h_{u \mu}) \right ) \cr
&& \phantom{m_{u\mu} \subst \frac{\partial}{\partial h_{u\mu}}\ln}
(\mu=1,2,\ldots,N); \cr
&& \bh_x  \subst \bH^{\rm T} \bm_u; \cr
&& \cu \subst -\frac{1}{N}\sum_{\mu=1}^N 
\frac{\partial^2}{\partial h_{u\mu}^2}
\ln \left (\int Ds
P(y_\mu|\sqrt{\hcu}s+h_{u \mu}) \right ); \cr
&& \Lambda_u \subst \frac{1}{\cu}-\hcu;
\nonumber
\end{eqnarray}
$\}$\\
{\bf {V-Step}}$\{$
\begin{eqnarray}
&&\mbox{Search $(\chi_x,\Lambda_x)$ for given $(\cu,\Lambda_u)$ 
to satisfy conditions} \cr
&& 
\quad \chi_x\!=\!\left \langle \frac{\Lambda_u}{\Lambda_x \Lambda_u+\lambda} 
\right \rangle_\rho \mbox{and }
\cu\!=\!\frac{1-\beta}{\Lambda_u}
\!+\!\left \langle \frac{\beta \Lambda_x}{\Lambda_x \Lambda_u+\lambda} 
\right \rangle_\rho; \cr
&&\hcw\subst \frac{1}{\chi_x}-\Lambda_x; \cr
&& \bh_x \subst \bh_x+\widehat{\chi}_x \bm_x; \cr
&& m_{x k} \subst \frac{\partial}{\partial h_{xk}}
\ln \left (\mathop{\rm Tr}_{x}
P(x) e^{-\frac{1}{2}\widehat{\chi}_x x^2+h_{xk} x} \right ) \cr
&& \phantom{m_{x k} \subst \frac{\partial}{\partial h_{xk}}\ln}
(k=1,2,\ldots,K); \cr
&& \bh_u  \subst \bH \bm_x; \cr
&& \chi_x \subst 
\frac{1}{K}\sum_{k=1}^K \frac{\partial^2}{\partial h_{xk}^2}
\ln \left (\mathop{\rm Tr}_{x} 
P(x) e^{-\frac{1}{2}\widehat{\chi}_x x^2+h_{x k} x} \right ); \cr
&& \Lambda_x \subst \frac{1}{\chi_x}-\widehat{\chi}_x;
\nonumber
\end{eqnarray}
$\}$
%\end{quote}
\normalsize
\caption{Pseudocode of the message-passing algorithm {\bf MPforPerceptron} 
\cite{ShinzatoKabashima2007}. The symbols 
``;'' and ``$\leftarrow$'' represent 
the end of a command line and the operation of substitution, respectively. 
The quantities $\Lambda_x$ and $\Lambda_u$ are the counterparts of 
$\Lambda_\xi$ and $\Lambda_\eta$ in Eq. (\ref{F_func})
for $\xi=\chi_x$ and $\eta=\chi_u$, respectively. 
}
\label{fig1}
\end{figure}

Unfortunately, the evaluation of Eq. (\ref{gen_gibbs_FE}) is 
also computationally difficult. 
One approach to overcoming this difficulty is to perform 
a Taylor expansion around $l=0$, for which 
$\widetilde{\Phi}(\bm_x,\bm_u;l)$ can be analytically 
calculated as an exceptional case, and substitute 
$l=1$ in the expression obtained
%, which is sometimes %referred to as the Plefka expansion 
\cite{Plefka1982}. 
However, the evaluation of higher-order terms, 
which are not negligible in general, 
requires a complicated calculation in this expansion, 
which sometimes prevents the scheme from being practically feasible. 
In order to avoid such difficulty, 
we take an alternative approach here, 
which is inspired by a derivative of Eq. (\ref{gen_gibbs_FE}), 
\begin{eqnarray}
\frac{\partial \widetilde{\Phi}(\bm_x,\bm_u;l) }{
\partial l}=-\left \langle ({\rm i} \bu)^T \bH \bx \right \rangle_l, 
\label{internal_energy}
\end{eqnarray}
following a strategy proposed by Opper and Winther \cite{OpperWinther2005}. 
Here, $\left \langle \cdots \right \rangle_l$ represents the 
average with respect to the generalized weight 
$\prod_{\mu=1}^N \widehat{P}(y_\mu|u_\mu)\times$
$\prod_{k=1}^K P(x_k)\times$ $\exp \left (
\bh_x \cdot \bx+\bh_u \cdot ({\rm i}\bu)+
({\rm i} \bu)^{\rm T}(l\bH) \bx \right )$, 
in which $\bh_x$ and $\bh_u$ are determined so as to 
satisfy $\left \langle \bx \right \rangle_l=\bm_x$ and
$\left \langle ({\rm i}\bu) \right \rangle_l=\bm_u$, 
respectively. 
The right-hand side of this equation is an 
average of a quadratic form composed of many random variables. 
The central limit theorem 
implies that such an average does not depend on the details 
of the objective distribution, but is determined 
only by the values of the first and second moments. 
In order to construct a simple approximation scheme, 
let us assume that the second moments are
characterized macroscopically by 
$\left \langle |\bx|^2 \right \rangle_l-
|\left \langle \bx \right \rangle_l|^2 =K \chi_x$
and $\left \langle |\bu|^2 \right \rangle_l-
|\left \langle \bu \right \rangle_l|^2 =N \cu$. 
Evaluating the right-hand side of Eq. (\ref{internal_energy}) using 
a Gaussian distribution, 
the first and second moments of which are 
constrained to be 
identical to those of the generalized weight, 
and integrating from $l=0$ to $l=1$, we have
\begin{eqnarray}
&&\widetilde{\Phi}(\chi_x,\cu,\bm_x,\bm_u;1)-
\widetilde{\Phi}(\chi_x,\cu,\bm_x,\bm_u;0) \cr
&& \simeq -\bm_u^{\rm T} \bH \bm_x -N F(\chi_x,\cu), 
\label{AdaTAP}
\end{eqnarray}
where the function $F(\xi,\eta)$ is provided as in 
Eq. (\ref{F_func}) by the empirical eigenvalue 
spectrum of $\bH^{\rm T}\bH$, $\rho(\lambda)=K^{-1}
\sum_{k=1}^K \delta(\lambda-\lambda_k)$ and the 
macroscopic second moments $\chi_x$ and $\cu$ are included 
in arguments of the Gibbs free energy because the right-hand side 
of Eq. (\ref{internal_energy}) depends on these moments. 
Eq. (\ref{AdaTAP}) offers a computationally feasible 
approximation of Eq. (\ref{gen_gibbs_FE}) for $l=1$, 
since assessment of $\widetilde{\Phi}(\chi_x,\cu,\bm_x,\bm_u;0)$, 
in which one can perform summations with respect to 
relevant variables independently, 
can be achieved at a reasonable computational cost.  

Although evaluation of Eq. (\ref{AdaTAP}) is computationally 
feasible, searching for saddle points of this function 
within a practical time is still a non-trivial problem. 
In Fig. \ref{fig1}, we present a message-passing type algorithm, 
which was recently proposed for a classification problem of single layer 
perceptrons \cite{ShinzatoKabashima2007}, 
as a promising heuristic solution for this problem. 
 
The efficacy of this method under appropriate conditions was experimentally 
confirmed for the perceptron problem, and to the extent to which it has been
 applied to several ensembles of linear vector channels, 
this algorithm has also been shown to exhibit a reasonable performance 
for the current inference task as well. However, 
its properties including convergence 
conditions have not yet been fully clarified,  and, 
therefore further investigation is necessary
for the theoretical validation and improvement 
of the performance of this method.

\section{Example: Welch bound equality sequences}
In order to demonstrate the utility of the proposed approach, 
let us apply the current methodologies to the analysis 
of the matrix ensemble that is characterized
by $\rho(\lambda)=(1-\beta^{-1})\delta(\lambda)
+\beta^{-1} \delta(\lambda -\beta)$ under the assumption $\beta > 1$, 
which corresponds to the case of Welch bound 
equality sequences (WBES) \cite{Welch1974}. 
We focus on the case of the Gaussian channel $P(y|\Delta)=(2 \pi \sigma^2 )^{-1/2} 
\exp \left (-(y-\Delta)^2/(2 \sigma^2) \right )$
and binary inputs $\bx \in \{+1,-1\}^K$, since this constitutes a simple, yet non-trivial problem.   
Under these assumptions, the developed framework has a higher capability than is required for   the assessment of the typical communication performance
with respect to the matrix ensemble, which can be carried out 
by a simpler method developed by the author and his colleagues 
\cite{TakedaUdaKabashima2006,TakedaHatabuKabashima2007}, as
was recently shown by Kitagawa and Tanaka \cite{KitagawaTanaka2007}. 
Nevertheless, the framework is still useful 
as one can derive a computationally feasible approximate 
decoding algorithm of good convergence properties based on 
the procedure shown in Fig. \ref{fig1}.

For Gaussian channels, $\Lambda_u$ in Fig. \ref{fig1} 
can be fixed as $\Lambda_u=\sigma^2$ in general. 
This yields an algorithm 
\begin{eqnarray}
\bm_u^{t+1}&=&\frac{1}{\sigma^2+\widehat{\chi}_u^t}
\left (
\by-\bH\bm_x^t+\widehat{\chi}_u^t\bm_u^t \right ), \label{Hstep} \\
m_{xk}^{t+1}&=&\tanh \left (\sum_{\mu=1}^N H_{\mu k} m_{u \mu}^{t+1}
+\widehat{\chi}_x^{t+1}m_{xk}^{t} \right ) \label{Vstep} \\
&&\phantom{aaaaaaaaaaa}(k=1,2,\ldots,K), \nonumber
\end{eqnarray}
for WBES, where $t$ denotes the number of iterations. 
$\widehat{\chi}_u^t$ in Eq. (\ref{Hstep}) is 
provided as $\widehat{\chi}_u^t=\beta/\Lambda_x^t$, 
where $\Lambda_x^t$ is determined so as to satisfy
$\chi_x^t=(1-\beta^{-1})/\Lambda_x^t+\beta^{-1}
\sigma^2/(\sigma^2 \Lambda_x^t +\beta)$ for given 
$\chi_x^t=1-K^{-1}|\bm_x^t|^2$. 
Utilizing the identical $\Lambda_x^t$, 
$\widehat{\chi}_x^{t+1}$ in Eq. (\ref{Vstep}) is 
evaluated as $\widehat{\chi}_x^{t+1}=1/\chi_x^t-\Lambda_x^t$. 

\begin{figure}[t]
\setlength{\unitlength}{1mm}
\begin{picture}(100,60)
\put(0,0){\includegraphics[width=80mm]{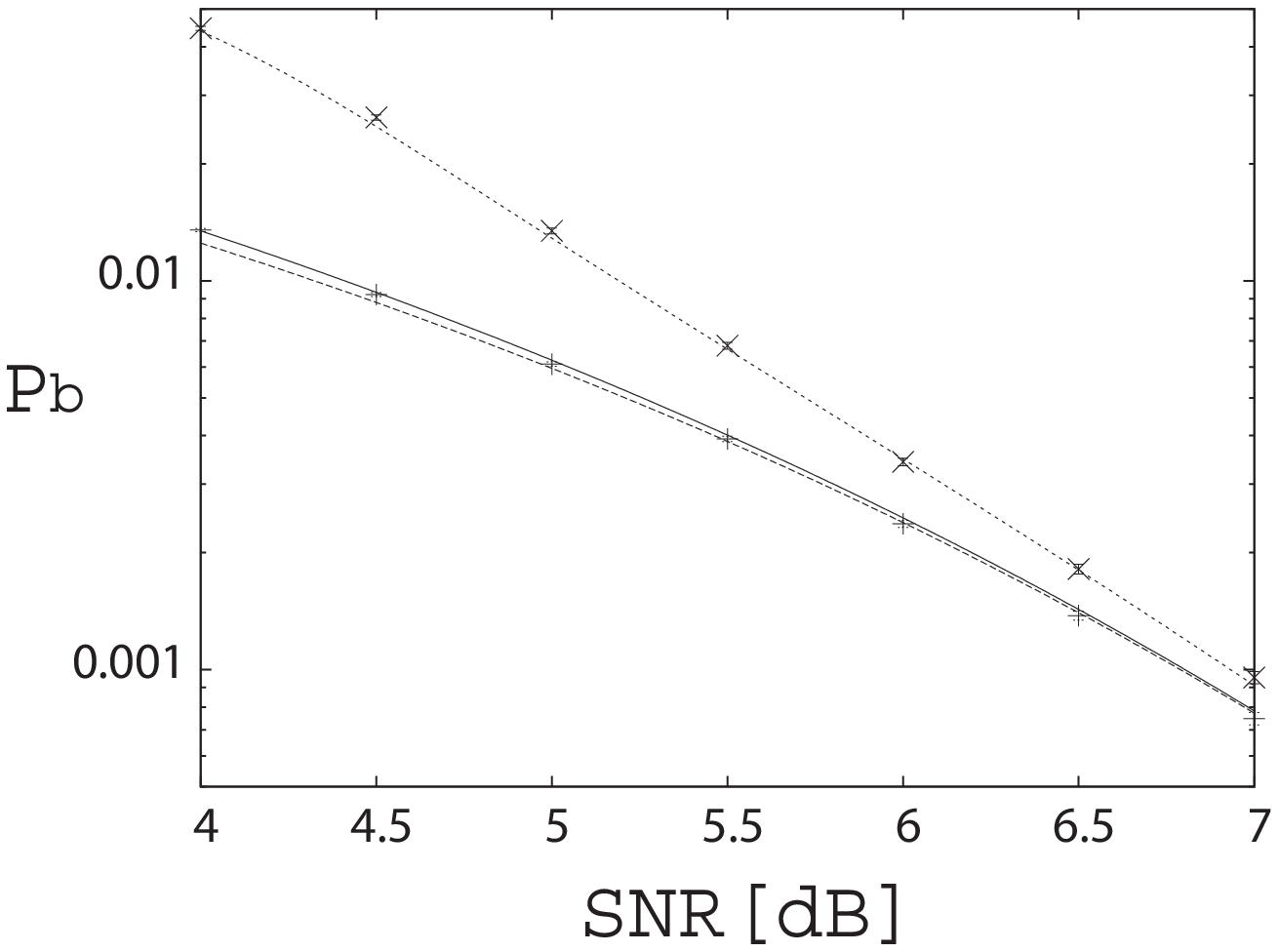}}
\end{picture}
\caption{BER vs. signal-to-noise ratio (SNR) for 
binary inputs for the case $\beta=1.1$.
The SNR plotted on the horizontal axis is given by 
$-10 \log_{10}(2 \sigma^2)$ while the vertical axis 
denotes the BER. The curves indicate theoretical predictions, 
which correspond to the scalar Gaussian channel, 
WBES and the basic matrix ensemble (BASIC) from the bottom. 
Sample matrices of BASIC are composed of IID entries of 
zero mean and $1/N$ variance Gaussian random variables. 
Values for WBES and BASIC are assessed by the replica method. 
The markers indicate experimental estimates of 
the BER obtained from 500 sample systems with $K=2048$ and $N=1862$
% by means of 
on the basis of the algorithm shown in Fig. \ref{fig1}. 
Excellent agreement between the curves and markers
validates both the performance analysis based on the replica method
and that of the developed approximation algorithm. 
}
\label{fig2}
\end{figure}

\begin{figure}[t]
\setlength{\unitlength}{1mm}
\begin{picture}(100,60)
\put(0,0){\includegraphics[width=80mm]{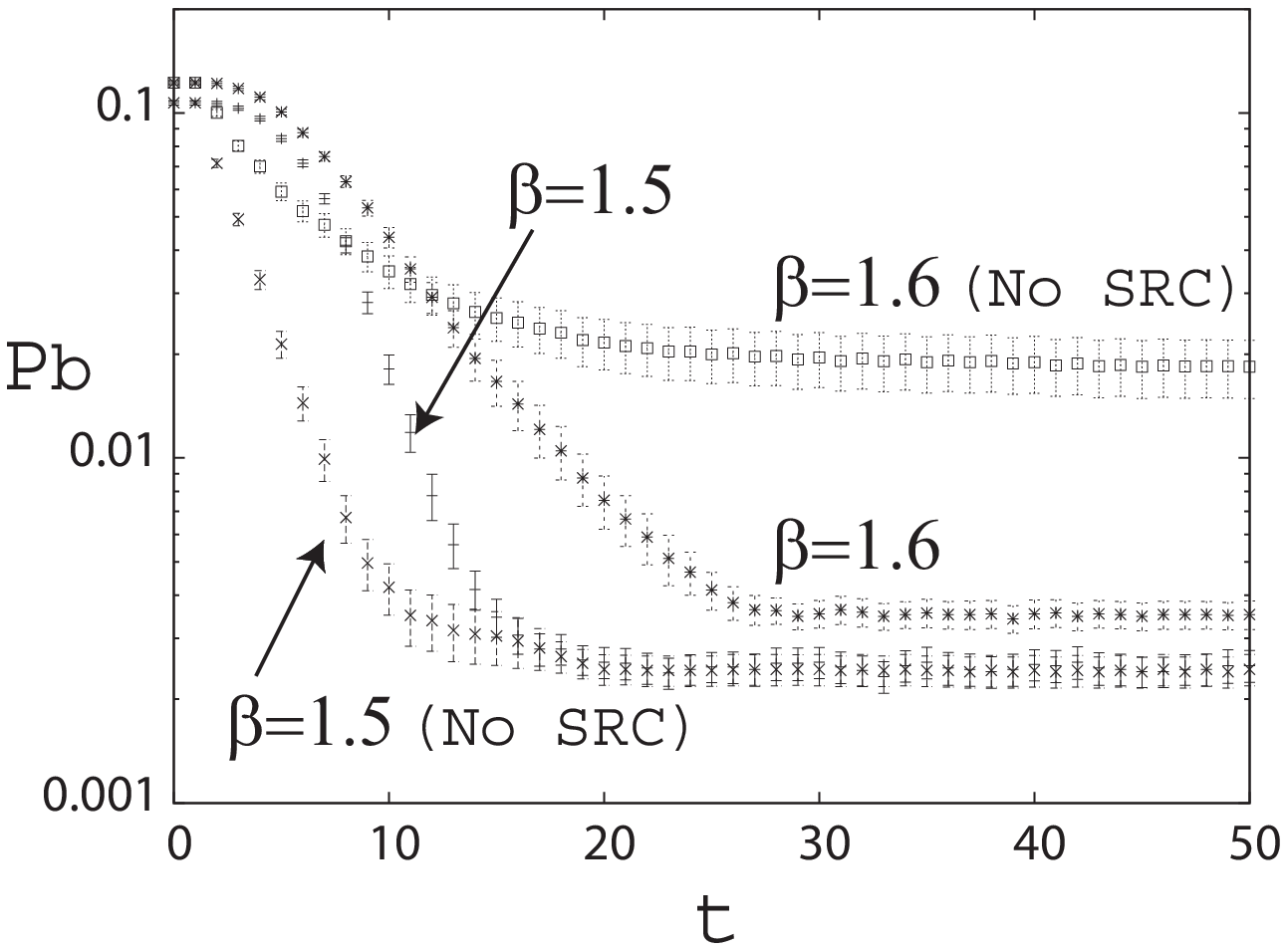}}
\end{picture}
\caption{
Influence of the cancellation of self-reactions 
in the approximate decoding for WBES. 
Experimentally assessed trajectories are plotted for two 
approximation algorithms for the cases $\beta=1.5$ and $1.6$, 
where SNR is fixed to 6.0. 
The horizontal axis represents 
the number of iterations, while the vertical axis denotes the BER. 
The data are obtained from 100 experiments of $K=512$ systems. 
The first algorithm used in these experiments is that presented by Eqs. (\ref{Hstep}) and 
(\ref{Vstep}) while the second algorithm, the results of which are denoted by 
``No SRC'' in the figure, is found by considering vanishing values of the macroscopic variables $\widehat{\chi}_u^t$ 
and $\widehat{\chi}_x^{t+1}$ in Eqs. (\ref{Hstep}) and (\ref{Vstep}). 
For both algorithms, initial states in the experiments were
set as $m_{xk}=\tanh(\theta_k/\sigma^2)$, 
where, in contrast to Fig. \ref{fig1}, $\theta_k=\sum_{\mu=1}^N H_{\mu k}y_\mu$
$(k=1,2,\ldots,K)$.  
For $\beta=1.5$, both algorithms 
converge to almost identical solutions, although the
 convergence of the first algorithm is slower. 
However, for $\beta=1.6$, the second algorithm converges to 
solutions of significantly higher BER while solutions found by 
the first algorithm still exhibit relatively low values of the BER. 
This indicates that the cancellation of the self-reaction terms 
in the approximation algorithm 
 acts to maintain the quality of the convergent solutions 
 for larger values of $\beta$. 
}
\label{fig3}
\end{figure}

Fig. \ref{fig2}  compares the BER for the  
theoretical assessment by the replica method with the 
experimental evaluation obtained by the algorithm 
of Eqs. (\ref{Hstep}) and (\ref{Vstep}). 
In the experiments, the estimates of the binary messages
are computed as $\widehat{x}_k={\rm sign}(m_{xk})$
for $k=1,2,\ldots, K$, where ${\rm sign}(a)=a/|a|$ for $a \ne 0$. 
This decoding scheme is optimal for minimizing BER
if $\bm_x$ represents the correct posterior 
average (\ref{postav}) \cite{Iba1999}. 
The excellent agreement between 
the curves and markers in this plot validates both the performance 
assessment based on the replica method and that based 
on the developed algorithm. 
A characteristic feature of Eqs. (\ref{Hstep}) and (\ref{Vstep})
is the inclusion of macroscopic variables 
$\widehat{\chi}_u^t$ and $\widehat{\chi}_x^{t+1}$, 
which are expected to act to cancel the self-reactions 
from previous states. \cite{TAP1978}. 
Fig. \ref{fig3} plots the influence of this operation, 
indicating that the cancellation acts 
to maintain the quality of the converged solution up 
to larger %values of 
$\beta$ under a condition of fixed SNR. 

\section{Summary}
In summary, we have developed a framework to analyze  linear
vector channels in a large system limit. 
The framework is based on the assumptions 
that the left and right singular value bases of the channel
matrix can be regarded as independently drawn from Haar 
measures over orthogonal (unitary, if the number field is defined 
over the complex variables) groups, 
and that the eigenvalues of the cross correlation matrix of the channel 
matrix asymptotically approach a certain specific distribution in the limit of large matrix size. 
These modeling assumptions allow a  characterization of the system 
in terms of an integral formula in two variables, which is fully determined by 
the eigenvalue distribution. Upon applying this 
formula in conjunction with the replica method, 
we have derived a general expression for 
the typical mutual information of general memory-less channels
with factorizable priors of continuous/discrete inputs.  
We have further proposed 
a computationally feasible decoding algorithm based on the formula, 
and have found that numerical results obtained from this 
algorithm are in excellent agreement with the theoretical predictions 
evaluated by the replica method. 

Future research directions include the application of 
the developed framework to various models of 
linear vector channels, and further improvement of 
the computationally feasible decoding algorithm.

\section*{Acknowledgments}
The author thanks Jean-Bernard Zuber 
for useful discussions concerning Eq. 
(\ref{F_func}). This research was supported in part 
by Grants-in-Aid MEXT/JSPS, Japan, Nos. 17340116 and 18079006.



%\bibliographystyle{IEEEtran}

\end{document}